\documentclass{aa}
\usepackage{graphicx}
\usepackage{natbib}
\begin{document}
   \title{Turbulence and particle acceleration in collisionless
	supernovae remnant shocks} 
   \subtitle{I-Anisotropic spectra solutions}

   \author{G.Pelletier
          \inst{1}
          \and
	   M.Lemoine \inst{2}          
	   \and
	   A.Marcowith \inst{3}
          }

   \offprints{A.Marcowith}

   \institute{Laboratoire d'Astrophysique de Grenoble, \\
	CNRS, Universit\'e Joseph Fourier II, \\
	BP 53, F-38041 Grenoble, France; \\
              and Institut Universitaire de France\\
              \email{Guy.Pelletier@obs.ujf-grenoble.fr}
         \and
              Institut d'Astrophysique de Paris, \\ 
	UMR 7095 CNRS, Universit\'e Pierre \& Marie Curie, \\
	98 bis boulevard Arago, F-75014 Paris, France\\
              \email{lemoine@iap.fr}
	 \and
              Centre d'\'Etudes Spatiales et du Rayonnement, \\ 
	CNRS, Universit\'e Paul Sabatier,\\
 	9, avenue du Colonel Roche, F-31028 Toulouse C\'edex, France\\
	\email{Alexandre.Marcowith@cesr.fr}
}

   \date{Received - ; accepted -}

\abstract{This paper investigates the nature of the MHD turbulence
excited by the streaming of accelerated cosmic rays in a shock wave
precursor. The two recognised regimes (non-resonant and resonant) of
the streaming instability are taken into account. We show that the
non-resonant instability is very efficient and saturates through a
balance between its growth and non-linear transfer. The cosmic-ray
resonant instability then takes over and is quenched by advection
through the shock. The level of turbulence is determined by the
non-resonant regime if the shock velocity $V_{\rm sh}$ is larger than
a few times $\xi_{\rm CR} \ c$, where $\xi_{\rm CR}$ is the ratio of
the cosmic-ray pressure to the shock kinetic energy. The instability
determines the dependence of the spectrum with respect to
$k_\parallel$ (wavenumbers along the shock normal). The transverse
cascade of Alfv\'en waves simultaneously determines the dependence in
$k_{\perp}$. We also study the redistribution of turbulent energy
between forward and backward waves, which occurs through the
interaction of two Alfv\'en and one slow magneto-sonic
wave. Eventually the spectra at the longest wavelengths are found
almost proportional to $k_{\parallel}^{-1}$. Downstream, anisotropy is
further enhanced through the compression at shock crossing.
 } \maketitle

\section{Introduction}
Fermi acceleration of cosmic rays in astrophysical shock fronts
depends in a crucial way on their transport properties in the
turbulent magnetic field on both sides of the shock. Often the
turbulent field spectrum and intensity are arbitrarily prescribed,
assuming that it has been built by the ambient medium independently of
the shock acceleration process. However \citet{LER}, \citet{WEN} have
argued for a long time that the development of an anisotropy of the
cosmic ray distribution function triggers an instability upstream of
the shock. \citet{VK} have investigated the consequences of this
phenomenon in the energy budget of the shock, in particular, with
respect to the efficiencies of conversion of the kinetic energy into
thermal, turbulent magnetic and cosmic ray energies. Recently,
\citet{BL} have shown that the amplification of the turbulent magnetic
energy could be quite significant, producing a magnetic field
intensity suitable to push the high energy cut-off of the proton
distribution up to the ``knee'' of the cosmic ray spectrum ($E\sim 3
\times 10^{15}\,$eV).\\ This theory has then been developed further
with accurate investigations of supernovae remnants (SNR). In
particular \citet{PT} have analysed the generation of turbulence and
emphasized the importance of advection for the saturation of the
spectrum.  They have also carried out a preliminary examination of the
role of the Kolmogorov cascade in the energy transfer among the
excited waves. More recently \citet{AB} described a non-resonant
regime of the streaming instability and has shown that its growth rate
should be dominant in the high wavenumber range (to be discussed in
more detail below). The fast growth of non-resonant modes could
provide the necessary magnetic field intensity during the early stages
of the SNR evolution to accelerate cosmic rays even up to the
cosmic-ray spectrum ``ankle'' at $E \sim 3 \times 10^{18}\,$ eV. If
verified this possibility would bring a strong support to the standard
galactic cosmic-ray model [see the discussion in \citet{Druryetal01}
and references therein].

In this paper, we analyze the excitation of Alfv\'en waves as a
function of the location in the upstream flow and of the wavenumber
taking into account the two instability regimes (Section 2). In
Sections 3 and 4, we calculate the saturation mechanism of the
instability considering the advection effect as a function of the
wavenumber and the location in the upstream flow. We calculate the
contributions of two non-linear effects: the transverse non-linear
transfer among turbulent Alfv\'en waves, and the non-linear
backscattering of Alfv\'en waves off slow magneto-sonic waves. These
two processes are shown to be relevant and essential to the
determination of the anisotropic turbulence spectra. We finally derive
these spectra which are essential to calculate the cosmic-rays
transport coefficients. The detailed calculation of these transport
coefficients is carried out in the companion paper \citep{MPL05}
, hereafter paper II. In section 5, we examine the consequences in the
downstream flow of the upstream excitation of the turbulence. In
particular we propose a spectrum for the turbulent magnetic field and
estimates of the relaxation length and of the parameters describing
the dynamo action downstream. The technical derivations are presented
in two appendices.

\section{Upstream excitation of MHD turbulence}
The instability triggered by the super-Alfv\'enic flow of cosmic rays
upstream of a shock has been analyzed in two ways: one is related to
the resonant interaction of the cosmic rays with the Alfv\'en waves
\citep{VK} and is essentially described by a kinetic theory.  The
other one has been recently proposed by \citet{AB} and emphasized
the importance of non-resonant interactions, in which the DC-electric
current of cosmic-rays generates a Lorentz force responsible for the
amplification of the MHD perturbations. In fact, this is the return
current in the background plasma which generates the perturbations
under some conditions. Both resonant and non-resonant interactions
are actually two regimes of the {\it same} streaming instability.

\subsection{The non-resonant regime of the instability}
The idea developed by \citet{AB} states that the cosmic-ray fluid
weakly responds to perturbations of wavelengths shorter than their
Larmor radii, so that the main response is in the form of a perturbed
current of the background plasma. Actually, the major role played by
the cosmic rays is to generate a DC return current in the
plasma. Because there are resonant interactions with cosmic-rays of
Larmor radius $r_{\rm L}$ at MHD scales $k$ such that $kr_{\rm
L}(\epsilon)= 1$, the validity criteria for the dominance of
non-resonant interactions needs to be analysed carefully. Indeed, if
one states that it holds for wavelengths shorter than the shortest
CR-Larmor radius, then one has to pay attention to the possibility of
going beyond the validity of MHD description, requiring $k\delta_0 < 1$
where $\delta_0 \equiv V_A/\omega_{\rm ci}$.\\

Let us first reformulate the calculation performed by \citet{AB} as
follows. The plasma remains locally neutral, so that the electric
charge carried by the cosmic rays is balanced by an electric charge
carried by the background plasma; $n_0$ being the number density of
electrons or protons, the number density of protons in the cosmic ray
component is $\chi_p n_0$ ($\chi_p < 1$) and the electrons contribute
to the CR-population with a number density $\chi_e n_0$ ($\chi_e <
1$); the charge density in the CR-population is therefore $\rho_{\rm
CR} = (\chi_p-\chi_e)n_0e$. Similarly, the electric current generated
by the Fermi process upstream is balanced by an electric current in
the thermal plasma; thus we state that $\rho_{\rm CR} + \rho_{\rm pl}
= 0$ and $\vec J_{\rm CR} + \vec J_{\rm pl} = \vec 0$, and $J_{\rm CR}
= (\chi_p-\chi_e)n_0eV_{\rm sh}$ with respect to the upstream medium,
$V_{\rm sh}$ being the shock velocity. Therefore, the electron
component of the thermal plasma drifts with respect to the ion fluid
with velocity $\vec V_{\rm d} = \frac{\chi_p-\chi_e}{1-\chi_e}\vec
V_{\rm sh}$. This drift does not destabilise the slow magneto-sonic
waves as long as it remains smaller than the sound velocity $c_{\rm
s}$, which holds when the CR-population is sufficiently tenuous that
$(\chi_p-\chi_e)/(1-\chi_e) < {\cal M}^{-1}$ (${\cal M}=V_{\rm
sh}/c_{\rm s}$ is the shock Mach number). When the drift velocity
exceeds the instability threshold, the fast growth of magneto-sonic
modes generates an anomalous resistivity that diminishes the current
down to a value close to the threshold value. Or in a more explicit
way, the magneto-sonic waves are amplified at a rate proportional to
$V_{\rm d} - c_s$ when the drift velocity of the thermal electrons exceeds
the sound velocity.  The electrons are scattered by these waves
through Landau resonant interactions and thus undergo an effective
collision frequency proportional to the energy of the waves.  This
gives rises to an anomalous resistivity that tends to reduce the
electric current as the wave amplitude grows and thus to make the
drift velocity $V_d$ decrease to a value close to the sound velocity.

Hereafter, we assume for simplicity that the magnetic field lies along
the shock normal. The generalization of our results to oblique
situations is straightforward, as long as a de Hofmann-Teller
transformation is possible.

In the non-resonant regime of the streaming instability, the
cosmic-ray fluid is, in a first approximation, passive: only the
thermal plasma responds while experiencing a Lorentz force due to its
charge $\rho_{\rm pl}$ and its current $\vec J_{\rm pl}$, under the
frozen in condition of the magnetic field, namely $\vec E+\vec u
\times \vec B = \vec 0$; where $\vec u$ is the local fluid
velocity. \\ The Alfv\'en wave equation is then modified as follows:
\begin{equation}
\label{Eq:Alfven}
\frac{\partial^2}{\partial t^2}\vec u -V_A^2
\frac{\partial^2}{\partial z^2}\vec u = \rho_{\rm pl} \frac{\vec
B_o}{\rho_0} \times(\frac{\partial}{\partial t}\vec u+V_{\rm sh}
\frac{\partial}{\partial z}\vec u) \ .
\end{equation}
This leads to two dispersion relations for right and left modes, namely:
\begin{equation}
\label{eq:disp}
\omega^2-k_{\parallel}^2V_{\rm A}^2 \pm k_{\rm c}V_{\rm A}^2(k_{\parallel} -
\frac{\omega}{V_{\rm sh}}) = 0 \ ,
\end{equation}
where 
%\begin{equation}
%\label{Eq:kc}
%k_{\rm c} \equiv \frac{\vert \rho_{\rm pl}B_o\vert}{\rho_0V_{\rm A}^2} = 
%\vert \chi_p-\chi_e\vert \frac{eB_o}{m_pV_{\rm A}}\frac{V_{\rm sh}}{V_{\rm A}} \ ,
%\end{equation}
\begin{equation}
\label{Eq:kc}
k_{\rm c} \equiv \frac{\vert \rho_{\rm pl}B_o V_{\rm sh}\vert}{\rho_0V_{\rm A}^2} = 
\vert \chi_p-\chi_e\vert \frac{eB_o}{m_pV_{\rm A}}\frac{V_{\rm sh}}{V_{\rm A}} \ , 
\end{equation}
and $\rho_0 = m_p \ n_0$.\\
The magnetic field $B_o$ and the Alfv\'en velocity $V_{\rm A} =
B_o/\sqrt{\rho_0 \ \mu_0}$ are to be considered as mean values.  The
scale $k_{\rm c}^{-1}$ must be compared to the minimum MHD scale
$\delta_0 \equiv V_{\rm A}/\omega_{\rm ci} = c/\omega_{\rm pi}$, below
which MHD no longer applies, and one gets $k_{\rm c}\delta_0 = \vert
\chi_p-\chi_e\vert V_{\rm sh}/V_{\rm A}$.  From the previous
dispersion relations, we easily deduce that the waves are stable when
$V_{\rm sh} < V_{\rm A}$, and that they become unstable only when
$V_{\rm sh}$ is sufficiently larger than $V_{\rm A}$ and
$k_{\parallel} < k_{\rm c}$. Indeed for $V_{\rm A}/V_{\rm sh} \ll 1$,
one of the branch is unstable when $\omega^2 = - V_{\rm
A}^2(k_{\parallel}k_{\rm c} - k_{\parallel}^2) < 0$. This unstable
mode is of right or left circular polarization depending on the main
composition of the CR-fluid and the orientation of the magnetic
field. Let $\vec b$ be the unitary vector that points toward the
same direction as the vector $-(\chi_p-\chi_e)\vec B_o$; then one gets
${\vec u} = i \vec b \times \vec u$: the mode is thus of right
circular polarization with respect to the direction defined by $\vec
b$. For the likely case of a proton dominated CR-fluid, the mode is
left-handed with respect to $\vec B_o$; in other words, it rotates in
the same sense as the protons. Such a left mode exists only for
$k\delta_0 \ll 1$, otherwise it is heavily damped by resonant
cyclotron absorption.
\bigskip

%The condition $k_c \delta_0 < 1$ is likely fulfilled; but $k_c$ could
%be at a scale for which this analysis does not apply because of the
%importance of response of the CR-fluid. 

\noindent \underline{Modification of the instability due to the
CR-response}\ :\\ Following Bell's formulation, the modification is
described by the complex factor $\sigma_*(k)$ such that
\begin{equation}
\label{Eq:disp}
\omega^2 - k_{\parallel}^2V_{\rm A}^2 \pm V_{\rm A}^2 k_{\rm
c}k_{\parallel}(1-\sigma_*) = 0 \ ,
\end{equation}
where corrections in $V_{\rm A}^2/V_{\rm sh}^2$ were neglected and
$\sigma_*$ reads:
\begin{equation}
\label{Eq:sigma_star}
\sigma_* \equiv \frac{1+\varepsilon}{\lambda_*^{1+\varepsilon}}
\int_0^{\lambda_*} \lambda^\varepsilon \sigma_p(\lambda)d\lambda \ .
\end{equation}
In this work, we assume a CR distribution function $f(p)\propto
p^{-4-\varepsilon}$ between $p_0$ and $p_{max}$, where $\varepsilon$
may be either positive or negative (see paper II, Marcowith {\it et
al.} 2005); moreover
\begin{equation}
\label{Eq:sigmap}
\sigma_p(\lambda) = \frac{4}{3}\lambda(1-\lambda^2)\left[\ln
\left(\frac{1+\lambda}{1-\lambda}\right) +i\pi\right] + \frac{3}{2}\lambda^2 \ ,
\end{equation}
with $\lambda \equiv [k_{\parallel} r_{\rm L}(p)]^{-1}$ and $\lambda_*
\equiv (k_{\parallel}r_*)^{-1}$ where $r_* \equiv r_{\rm
L}(p_*)$. Here $p_*$ ($r_*$) sets the minimum momentum (Larmor radius)
of the cosmic-ray distribution function.  This cut-off depends in
principle on the distance $x$ to the shock front (measured along the
shock normal), since the cosmic ray distribution function roughly
decreases as $f(p,x)\propto \exp[-x/\ell_{\rm D}(p)]$;
$\ell_{\rm D}(p)=(1/3)c\tau_{\rm s}(p)c/V_{\rm sh}$ is related to the
scattering time $\tau_{\rm s}(p)$ and is an increasing function of
$p$. Hence at each distance $x$ there exists $p_*(x)$, defined by $x =
\ell_{\rm D}(p_*)$, such that the contribution of the smaller energies
$p<p_*$ is negligible, since the corresponding diffusion lengths are
short. The cosmic-ray density is $n_{\rm CR} = \int_{p_0}^{p_{max}}
{\rm d}^3p f(p)$ is related to the CR pressure at the shock front via
$n_{\rm CR} = 3P_{\rm CR}/(\Phi p_0c)$, with $\Phi$ a dimensionless
number of order $\log(p_{\rm max}/m_pc)\sim 10$.\\

For short waves, $k\gg 1/r_*$ or $\lambda_* \ll 1$, $\Re e\{\sigma_*\}
\simeq 0$ and the previous result of Eq.~(\ref{eq:disp}) holds. In
particular, the non-resonant growth rate is 
\begin{equation}
\label{Eq:Gnr}
G_{\rm n-res}(k_\parallel)\simeq V_{\rm A}(k_{\rm c}k_\parallel)^{1/2}
\end{equation} 
for $1/r_* \ll k_\parallel \ll k_c$. Note that this non-resonant
instability is not operative for a wavenumber $k_\parallel$ at
distances $x \ll \ell_{\rm D}(r_{\rm L}=1/k_\parallel)$ since the
corresponding $r_*(x) \ll 1/k_\parallel$, $r_*$ being an increasing
function of $x$. The exact spatial dependence of the growth rate will
be specified further on. The cut-off wavenumber $k_c$ is redefined 
from the Eq.(\ref{Eq:kc}) by:
\begin{equation}
k_{\rm c} \,=\, {4\pi n_{\rm CR}e V_{\rm sh}\over
B_o}\,=\,{12\pi\over\Phi}{P_{\rm CR}\over B_o^2}{V_{\rm sh}\over
c}{1\over r_*}\ .
\label{eq:kc}
\end{equation}

%\begin{equation}
%k_{\rm c} \,=\, {4\pi n_{\rm CR}e V_{\rm sh}\over
%B_o}\,=\,{12\pi\over\Phi}{P_{\rm CR}\over B_o^2}{V_{\rm sh}\over
%c}{1\over r_*}\ .
%\label{eq:kc}
%\end{equation}

For long waves, $k\ll 1/r_*$ or $\lambda_* \gg 1$, $\Re e\{\sigma_*\}
= 1$ and the CR-response dominates, i.e. the non-resonant instability
is inactive.

\subsection{The resonant regime of the instability}
The imaginary part of $\sigma_*$ describes the resonant interaction
between cosmic rays and Alfv\'en waves; it is responsible for a growth
rate that reaches a maximum for $\lambda_*=1$, and for longer waves
($\lambda_* > 1$), $\Im m\{\sigma_*\} =
\frac{3\pi}{16}\frac{1+\varepsilon}{\lambda_*^{1+\varepsilon}}$. However
we will adopt a slightly different description, in the sense that we
expect to get oblique Alfv\'en waves that essentially are of linear
polarisation, which changes the resonance conditions as both
electrons and ions, moving forward or backward, can resonate either
with forward modes or backward modes. The small instability growth
rate is the same, within an angular factor of order unity, and is
given by
\begin{equation}
\label{Eq:G_res}
G_{\rm res}(k_{\parallel},x) =
G_0(k_{\parallel},x)\phi(x/\ell_{\rm D}(k_{\parallel})),
\end{equation}
where $\ell_{\rm D}(k_\parallel)$ should be understood as $\ell_{\rm D}(r_{\rm
L}=1/k_\parallel)$; the exact spatial dependence of $\phi$ will be
specified further on. The growth rate $G_0$ is given by
\begin{equation}
\label{Eq:G0}
G_0(k_{\parallel},x) =
\frac{\pi}{4}\frac{\alpha_0(\varepsilon)}{\delta_0}\frac{n_{\rm
CR}}{n_0}\left(\frac{\cos \theta}{\vert \cos \theta \vert}V_{\rm sh} -
\frac{4+\varepsilon}{3}V_{\rm A}\right)(k_{\parallel}
r_*)^{1+\varepsilon} \ ,
\end{equation}
this expression can be found in \citet{MEL}, the coefficient
$$
\alpha_0(\varepsilon) = {1\over 2}(1+\varepsilon)(4+\varepsilon) \int
\vert \mu \vert^{1+\varepsilon}(1-\mu^2)\,{\rm d}\mu \ .
$$
Eq.(\ref{Eq:G0}) scales as $\overline{B}^{-(1+\varepsilon)}$ similarly
to \citet{PT}.  It clearly shows that only modes propagating forward
are destabilized when $V_{\rm sh}$ is sufficiently larger than the
Alfv\'en velocity. The resonant growth rate is maximum at the scale
$r_*$ and scales like $r_*^{1+\varepsilon}$ in regards to the distance
to shock front $x$. Note that the backward waves are damped at the
same rate than the forward waves are amplified. The original
calculation has been done by \citet{LER}, \citet{WEN} and used in the
theory of cosmic ray transport by \citet{SKI}, then by \citet{VK} for
the excitation of turbulence upstream of a shock. Hereafter we will
assume for simplicity $\varepsilon = 0$, the value of this parameter
will be discussed in paper II.

The function $\phi$ stems from the solution of the evolution equation
\begin{equation}
\label{Eq:EqDC}
V_{\rm sh} \frac{\partial}{\partial x} \phi + \frac{\partial}{\partial
x}D\frac{\partial}{\partial x} \phi = 0
\end{equation}
with $\phi(0) = 1$. This function $\phi$ represents the spatial
profile of the CR distribution function, which decreases with the
characteristic scale given by the diffusion length $\ell_{\rm D} \equiv
D/V_{\rm sh}$. The diffusion length depends on the Larmor radius
$r_{\rm L}$, and thus the instability growth rate $G(k_{\parallel})$
decreases with $\phi$ with a characteristic length which is the
diffusion length for $r_{\rm L} \simeq k_{\parallel}^{-1}$. This can
be derived rigorously from the general expression of the growth rate
that involves a resonance for $k_{\parallel}r_{\rm L} \mu = 1$ ($\mu$
is the particle pitch-angle cosine). In the case of uniform diffusion
coefficient $D$, $\phi = \exp(-x/\ell_{\rm D})$.

\section{Saturation mechanism and stationary spectra}

\subsection{The main elements of the theoretical description}

\noindent \underline{WKB-approximation}\ :\\ The turbulence spectrum
is not homogeneous but grows when approaching the shock front like the
cosmic ray distribution. The scale of spatial variation is given at
each energy by the diffusion length $\ell_{\rm D} \equiv (1/3)(c/V_{\rm
sh})c\tau_{\rm s}$, which is much larger than the Larmor radius at the
same energy; $\tau_{\rm s}>t_{\rm L}$ is the scattering time defined
further below as a function of the turbulence spectrum. Since the
mode, that undergoes a resonant interaction at this energy, has a
wavelength equal to $2\pi r_{\rm L}$, its wavelength is shorter than
the diffusion length and thus a WKB description of the evolution of
the spectrum is suitable.  This statement is also true for the
non-resonant modes.

\bigskip

\noindent \underline{Advection versus non-linear coupling}: \\ 

For each spatial scale (or each value of $k_{\parallel}$), there are
three relevant time scales: i) the growth time scales of the resonant
and non-resonant modes $G_{\rm res}(k_{\parallel},x)^{-1}$ and $G_{\rm
n-res}(k_{\parallel},x)^{-1}$, ii) the non-linear time scale that can
be defined as the eddy turn over time $\tau_{\rm n-lin}(k) = (k\bar
u(k))^{-1}$, where the turbulent velocity $\bar u(k)$ is such that
$\rho_0 \bar u(k)^2$ is the turbulent energy density at that scale, or
by an appropriate non-linear scattering time; iii) the advection time
$\tau_{\rm adv}(k)$ at which the mode is caught up by the shock front
that propagates faster than the forward waves ($V_{\rm sh} > V_{\rm
A}$).\\

The time scale $\tau_{\rm adv}(k_{\parallel}) =
\ell_{\rm D}(k_{\parallel})/V_{\rm sh}$ with $k_{\parallel}r_{\rm L} = 1$
and one obtains
\begin{equation}
\label{Eq:taua}
\tau_{\rm adv} = {1\over 3}{c^2\over V_{\rm sh}^2} \tau_{\rm
s}(1/k_{\parallel}) \ .
\end{equation}
The pitch angle frequency $\nu_{\rm s} = \tau_{\rm s}^{-1}$ is known
for an isotropic spectrum $S(k)$, and, if furthermore the spectrum is
a power law of the form $\eta k_{\rm min}^{-1} (k/k_{\rm
min})^{-\beta}$, then
\begin{equation}
\label{Eq:NUS}
\nu_{\rm s} \simeq \pi (\beta -1)\omega_{\rm L} \eta \rho^{\beta-1} \ ,
\end{equation}
with the rigidity defined as $\rho \equiv k_{\rm min}r_{\rm L}$ [see
\citet{Cassetal02}]. In the above equation, the prefactor $\beta-1$
should be replaced by $\left[\log(k_{\rm max}/k_{\rm min})\right]^{-1}$
if $\beta=1$. The turbulence level $\eta$ is defined in the next
paragraph.

However, we deal with anisotropic spectra of the form $S_{3\rm d}
\propto k_{\perp}^{-q} k_{\parallel}^{-\beta}$ with $q>2$, leading to
the same formula Eq.~(\ref{Eq:NUS}) (see paper II, for further details).
Hereafter, we will use 1D-spectra $S(k_{\parallel})$ defined such that

\begin{equation}
S_{3\rm d}(\vec k) \,=\, 2\pi (q-2)k_{\rm min}^{-2}
\left({k_{\perp}\over k_{\rm min}}\right)^{-q} S(k_{\parallel})\ ,
\end{equation} 
which implies notably  
\begin{equation}
\int {{\rm d}^3k\over (2\pi)^3} \ S_{\rm 3D}(k) = \int 
{{\rm d}k_\parallel\over 2\pi} \ S(k_\parallel)\ .
\end{equation}
For convenience, hereafter, the normalization of $S$ is defined in
regards to the magnetic energy density at the infinity
\begin{equation}
\label{Eq:normS}
\int \frac{{\rm d}k_\parallel}{2\pi} \ S(k_\parallel) \ = \ \int {\rm
d}\log(k_{\parallel}) \frac{\delta B^2(k_{\parallel})}{B_{\infty}^2} \
= \ \frac{\delta B^2}{B_\infty^2} \ ,
\end{equation}
where $\delta B$ is the turbulent field amplitude, and $B_\infty$ the
(original) uniform component to be taken in the interstellar medium
far upstream of the forward SN shock. The Alfv\'en velocity $V_{\rm
A\infty}$ in the interstellar medium can then be deduced
immediately. The magnetic field and Alfv\'en velocity amplified by the
streaming instability are hereafter noted $\overline{B}$ and
$\overline{V}_{\rm A}$.  The two Alfv\'en velocities are linked by
$\overline{V}_{\rm A} = V_{\rm A\infty}/(1-\eta)^{1/2}$ [see
\citet{PT}]. The quantity $\eta \equiv \delta B^2/(\delta B^2 +
B_\infty^2)$ determines the strength of turbulence; in particular
$\eta\rightarrow 1$ corresponds to $\delta B/B_\infty\rightarrow
+\infty$. Finally, the magnetic field turbulent amplitude at a scale
$k_{\parallel}$ and the 1D spectrum $S(k_{\parallel})$ are tied by the
relation: $\delta B^2(k_{\parallel}) = B^2_{\infty} \ k_{\parallel}
S(k_{\parallel})/(2\pi)$. 

While considering the resonant instability, we distinguish the forward
1D-spectrum $S^+$ of forward waves and the spectrum $S^-$ of backward
waves. The pitch angle frequency is the sum of the two contributions
because particles resonantly interact with both spectra irrespectively
of direction of motion (this is an important point related to the
resonance condition with linearly polarized Alfv\'en waves, as
mentioned before). The advection time is using the definition of
$\nu_{\rm s}$ in Eq.(\ref{Eq:NUS})
%\begin{equation}
%\label{Eq:TOA}
%\tau_{\rm adv}(k_{\parallel}) = {1\over 3}{c\over V_{\rm sh}^2}{1-\beta\over 2\pi}
%\frac{1}{k_{\parallel}^2}\frac{1}{S^++S^-}{1\over 1-\eta} \ .
%\end{equation}
\begin{equation}
\label{Eq:TOA}
\tau_{\rm adv}(k_{\parallel}) = {2\over 3}{c\over V_{\rm sh}^2}
\frac{1}{k_{\parallel}^2}\frac{1}{S^++S^-}{1\over 1-\eta} \ .
\end{equation}
Note that the advection time remains finite when the magnetic field
tends to be completely turbulent, i.e. when $\eta \rightarrow 1$. 

We introduce an important dimensionless quantity that measures the
ratio of advection time to instability growth time : $a(k_{\parallel})
\equiv 2G\tau_{\rm adv}$. For the resonant regime of the instability
using Eq.~(\ref{Eq:G0}) and Eq.~(\ref{Eq:TOA}) we obtain:
\begin{equation}
\label{Eq:a_res}
a_{\rm res}(k_{\parallel},x) = {\pi \over \Phi}{\cal M}_{\rm
A\infty}\xi_{\rm CR} {1\over k_\parallel(S^++S^-)}{\overline B \over B_\infty},
\end{equation}
%\begin{equation}
%\label{Eq:a_res}
%a_{\rm res}(k_{\parallel},x) = {\pi \over 2\Phi}{\cal M}_{\rm
%A\infty}\xi_{\rm CR} {1\over k_\parallel(S^++S^-)}{\overline B \over B_\infty},
%\end{equation}
where ${\cal M}_{\rm A\infty} = V_{\rm sh}/V_{\rm A\infty}$ is the
Alfv\'enic Mach number measured with respect to the interstellar
magnetic field value, $\xi_{\rm CR} \equiv P_{\rm CR}/\rho_0 V_{\rm
sh}^2$ is the ratio of CR to shock pressure and can reach values
as high as $0.5$ in non-linear acceleration models \citep{Betal96,BE99},
and $\overline B^2\equiv\delta B^2 + B_\infty^2$.

The ratio $\overline B/B_\infty$ stems from the spatial dependence of
$r_*$ in Eq.~(\ref{Eq:G0}) and from the $1-\eta$ factor in
Eq.~(\ref{Eq:TOA}); $\overline B/B_\infty$ depends only on the
distance to the shock front. The Eq.~(\ref{Eq:a_res}) accounts for the
amplification of the magnetic field along the normal to the shock
front and permits the inclusion of both non-resonant and resonant
regimes in the evolution equation (see Appendix A and B).  We define the
reference spectrum  $S_\star(k_\parallel)$
\begin{equation}
\label{eq:Sstar}
S_{\star}(k_\parallel) ={\pi \over \Phi}{\cal M}_{\rm A\infty}\xi_{\rm
CR} {1\over k_\parallel}{\overline B \over B_\infty} \ .
\end{equation}
%\begin{equation}
%\label{eq:Sstar}
%S_{\star}(k_\parallel) ={\pi \over 2\Phi}{\cal M}_{\rm
%A\infty}\xi_{\rm CR} {1\over k_\parallel}{\overline B \over B_\infty} 
%\end{equation}

Simultaneously the Alfv\'enic turbulence develops an energy transfer
mainly in the transverse direction which determines the shape of the
transverse spectrum in $k_{\perp}$. The non-linear transfer rate is
$t_{\rm n-lin}^{-1} \approx k_{\perp} \ \bar u_{\perp} \simeq k_{\perp} \
V_{\rm A \infty} \times [k_{\parallel} \ k_{\perp}^2 S_{3\rm
D}(k_{\parallel},k_{\perp})]^{1/2}$. We can define the efficiency of
the energy transfer process using Eq.~(\ref{Eq:normS}) and the
dimensionless number $\kappa_A$
%\begin{equation}
%\label{Eq:kappa_a}
%\kappa_A \equiv {\tau_{\rm adv} \over \tau_{\rm n-lin}} \sim
%\frac{c\overline{V}_{\rm A}}{V_{\rm sh}^2}{1\over \left[(1-\eta)\eta\right]^{1/2}}\ ,
%\end{equation}
\begin{equation}
\label{Eq:kappa_a}
\kappa_A \equiv {\tau_{\rm adv} \over \tau_{\rm n-lin}} \sim
\frac{c\overline{V}_{\rm A}}{V_{\rm sh}^2}\frac{1-\eta}{\eta}\ ,
\end{equation}

This number is sufficiently high for the Alfv\'enic cascade to fully
develop; this will be discussed in Section \ref{S:alfven_turb}.
Because the Alfv\'enic cascade does not convert energy from the
forward waves into backward waves, the backscattering of Alfv\'en
waves off slow magneto-sonic waves will also be considered and will
proved to be efficient to redistribute the energy over all the
spectra. This discussion is postponed to Section \ref{S:nlback}.

\subsection{The spatial profiles and spectra}

As mentioned earlier, the quantity $r_*(x)$ denotes the minimum Larmor
radius of streaming cosmic rays at a distance $x$ from the shock
front; $r_*(x)$ can be defined by the condition:
\begin{equation}
\int_0^x{{\rm d}x'\over \ell_{\rm D}[x',r_*(x')]}\,=\,1,
\label{eq:rstar}
\end{equation}
which, if $\ell_{\rm D}$ does not depend on $x$, amounts to
$x=\ell_{\rm D}[r_*(x)]$.

The non resonant regime of instability occurs for modes such that
$k_\parallel r_*(x)\gg 1$, hence at distances $x > x_{\rm
min}(k_\parallel)$ with $x_{\rm min}(k_\parallel)$ defined by
$r_*[x_{\rm min}(k_\parallel)]=1/k_\parallel$. Of course, this
non-resonant growth occurs provided there exists cosmic rays with
$r_{\rm L}>r_*(x)$, hence for $x <x_{\rm max}\sim \ell_{\rm D}(r_{\rm
L,max})$. In contrast, the resonant interaction growth rate is
maximal for $k_\parallel = 1/r_*(x)$, therefore the vicinity of the
shock front $x<x_{\rm min}(k_\parallel)$ is dominated by the resonant
regime of the instability.

\begin{figure}
\includegraphics[width=0.5\textwidth]{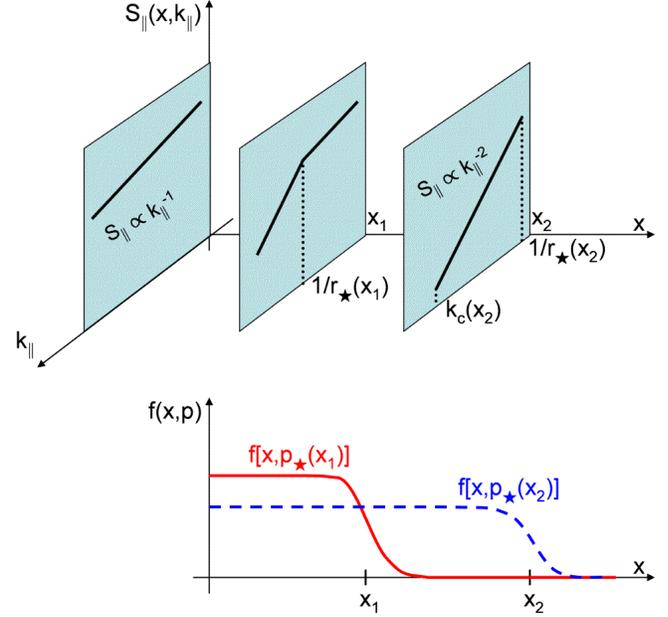} 
 \caption{Sketch of the evolution of the turbulence spectrum $S$ in the
 $(x,k_\parallel)$ plane (upper graph) and of the cosmic-ray
 distribution function $f(x,p)$ as a function of $x$ for two values of
 the momenta (lower graph). See text for details.}
\label{fig:spectra}
\end{figure}

In Fig.~(\ref{fig:spectra}) we sketch the simultaneous evolution of
the turbulence spectra as determined as a function of $x$ and
$k_\parallel$ and the distribution function of cosmic rays as a
function $x$ for two values of the momenta (hence two values of the
Larmor radius). At point $x_2$, far from the shock front located at
$x=0$, only the non-resonant instability has been active in the
momentum range $r_\star^{-1}(x_2)\leq k_\parallel\leq k_{\rm c}(x_2)$
and has produced a spectrum $S\propto k_\parallel^{-2}$ (Section
\ref{S:n-ressat}). At point $x_2$ as there are no cosmic rays with
Larmor radii such that the resonance condition $k_\parallel r_{\rm
L}\sim 1$ can be satisfied in the above momentum range, since
$r_\star(x_2)$ is by definition the minimum Larmor radius of cosmic
rays at point $x_2$. The lower graph of Fig.~(\ref{fig:spectra})
sketches accordingly the evolution of the distribution function of
cosmic-rays at momentum $p_\star(x_2)$, which corresponds to the
minimum momentum of cosmic-rays at $x_2$, or, equivalently, to a
Larmor radius $r_\star(x_2)$. The minimum momentum $p_\star$ (or
Larmor radius $r_\star$) is a growing function of $x$, in particular
$p_\star(x_2)>p_\star(x_1)$. Hence, at point $x_1$, the resonant
instability has been active in the momentum range $k_\parallel \la
1/r_\star(x_1)$ as there are cosmic-rays with Larmor radii that can
satisfy the resonance condition in this range. The corresponding
spectrum is $S\propto k_\parallel^{-1}$ (Section \ref{S:satr}). At
momenta $k_\parallel \ga 1/r_\star(x_1)$, only the non-resonant
instability has been active, for reasons similar to those discussed
above, and therefore the spectrum $S\propto
k_\parallel^{-2}$. Finally, at the shock front, the resonant
instability has overtaken the non-resonant instability over all the
wavelength range, so that the final spectrum $S\propto
k_\parallel^{-1}$.

% If the turbulence spectrum index $\beta$ differs at
%$y<1$ and $y>1$ due to the different nature of the instability in
%these regions, we notes that $y=\left(k_\parallel
%r_*\right)^{2-\beta_<}$ and $y_{\rm max}\equiv(k_\parallel
%r_{\rm L,max})^{2-\beta_>}$, where $\beta_<$ (resp. $\beta_>$) is that
%determined by the resonant (resp. non-resonant) instability for $y<1$
%(resp. $y>1$).

%In the attempt to take account of both regime of
%instability, $S_*$ must be modified, whereas it is in
%$k_{\parallel}^{-1}$ for $y<1$ because of the resonant regime, it is
%in $k_{\parallel}^{-3/2}$ for $y>1$ because of the Bell regime. Now
%this implies that for each $x$ the spectrum is broken. Indeed for
%$k_{\parallel} < 1/r_*(x)$, the spectrum is determined by the
%instability of the resonant modes, whereas for $k_{\parallel} >
%1/r_*(x)$ the spectrum is determined by the instability of the
%non-resonant modes.

%First, disregarding the non-linear transfer among waves, the evolution
%of the unstable spectrum is simply governed by
%\begin{equation}
%\label{ }
%\frac{\partial}{\partial y} S^+ = -a e^{-y} S^+ \ .
%\end{equation}

Since the two regimes of the instability take place in different
regions, we can solve for the spectra using in a first place the
equation involving solely non-resonant growth for $x_{\rm
min}(k_\parallel)<x<x_{\rm max}$ and then use this solution at $x_{\rm
min}(k_\parallel)$ as the initial condition for resonant growth up to
the shock front at $x=0$.

\subsubsection{The non-resonant regime}\label{S:n-ressat}

The equation governing the growth of the turbulent spectrum through
the non-resonant instability is, using Eq.(\ref{Eq:Gnr}):
\begin{equation}
V_{\rm sh}{\partial\over\partial
x}S(k_\parallel,x)\,=\,-2V_{A\infty}\sqrt{k_c(x)k_\parallel}\theta(x_{\rm
max}-x)S(k_\parallel,x)\ ,
\label{eq:non-res}
\end{equation}
where $k_c(x)$ is defined in Eq.~(\ref{eq:kc}). Assuming for the moment
that $\eta$ is constant ($=\eta_\infty$), Eq.~(\ref{eq:rstar})
then gives  $x=\ell_{\rm D}(r_*)$, or:
\begin{equation}
k_{\rm min}r_*(x)\,\simeq\, \left({3\eta_\infty V_{\rm sh}\over
c}\right)^{1/(2-\beta_<)}\left(k_{\rm min} x\right)^{1/(2-\beta_<)}\ ,
\label{eq:kr}
\end{equation}
where $\beta_<$ is the index of the turbulent spectrum in the vicinity
of the shock front, i.e. that which is produced by the resonant
instability.  Introducing
\begin{equation}
\zeta\,\equiv\, {2\over (3\eta_\infty)^{1-m}}{\cal M}_{\rm
A\infty}^{-1} \left({V_{\rm sh}\over c}\right)^{m-{1\over2}}
\left({12\pi\over\Phi}{P_{\rm CR}\over B_\infty^2}\right)^{1/2}\ ,
\label{eq:zeta}
\end{equation}
with $m = (3-2\beta_<)/(4-2\beta_<)$, then 
\begin{eqnarray}
S(k_\parallel,x) &=& S(k_\parallel,x_{\rm{max}}) \exp\Biggl\{-{\zeta
\over m} \sqrt{{k_\parallel} \over k_{\rm{min}}} \times \nonumber \\
& &\left[(k_{\rm{min}}x)^m - 
(k_{\rm{min}}x_{\rm{max}})^m \right]\Biggr\} \ .
\end{eqnarray}
The amplification factor is thus:
\begin{equation}
{S(k_\parallel,x=0) \over S(k_\parallel,x_{\rm{max}})} = \exp\left[{\zeta
\over m} \sqrt{{k_\parallel} \over k_{\rm{min}}}
(k_{\rm{min}}x_{\rm{max}})^m\right] \ .
\end{equation}
It turns out that $k_{\rm{min}}x_{\rm{max}} =
\ell_{\rm D}(r_{\rm{L,max}})/r_{\rm{L,max}} \gg 1$, and for $\beta_< = 1$
(corresponding to Bohm scaling, see next section) $\zeta \sim 1$.  The
amplification would be enormous, the instability would deplete the
shock quickly, unless another saturation mechanism occurred earlier.
Even if we accounted for a variation of $\eta$, the amplification level
would still be too large, as the following saturation mechanism keeps
the magnetic field energy density to a lower level.

\bigskip
\noindent{\underline{Saturation mechanism}:\\ 

In fact, the non-resonant growth should saturate much earlier. As the
magnetic field gets amplified beyond its initial value, one can
extrapolate the previous calculations by substituting $B_{<k}$ for
$B_{\infty}$, where $B_{<k}^2 = B_{\infty}^2 +
\int_{k_{\rm min}}^{k_{\parallel}} {\rm d}\log k'_{\parallel}\,\delta
B^2(k'_{\parallel})$ represents the average field on scales larger
than $k_{\parallel}^{-1}$, and again $\delta B(k_{\parallel})$ denotes the
amplified random component on scale $k_{\parallel}$. \\ 

For $\vert\chi_p-\chi_e\vert\sim\chi_p$, using Eq.~(\ref{Eq:kc}), the
cut-off wavenumber of the instability $k_{\rm c}$ can be written as
\begin{equation}
k_{\rm c} \,=\, {4\pi n_*e V_{\rm sh}\over
B_{<k}}\,=\,{12\pi\over\Phi}{P_{\rm CR}\over B_{<k}^2}{V_{\rm sh}\over
c}{1\over r_*},
\label{eq:kc}
\end{equation}
The non-resonant instability occurs for wavenumbers $1/r_* <
k_\parallel < k_{\rm c}$ and its saturation is achieved once $k_c r_*
= 1$.  This simple condition leads to a magnetic field energy density:
\begin{equation}
\label{Eq:Bsatnr}
\frac{B^2}{8\pi} \simeq \frac{3}{2\Phi} \frac{V_{sh}}{c} P_{CR} \sim
\frac{3\xi_{CR}}{2\Phi} \rho_0 \frac{V_{sh}^3}{c} \ .
\end{equation}
This last estimate is in agreement with Bell (2004). 

%can saturate at a scale $k_{\parallel}$
%either when $1/r_*=k_\parallel$ ($y=1$) or $k_\parallel=k_{\rm c}$,
%whichever comes first. The cut-off $k_{\rm c}$ depends on $y$ as
%$k_{\rm c}=(12\pi/\Phi)P_{\rm CR}(V_{\rm
%sh}/c)k_\parallel/(yB_{<k}^2)$, hence the saturation leads to a
%profile
%\begin{equation}
%S(k_\parallel,y)\,\sim\, \frac{2\pi \ B^2_{<k}}{K \ k_{\parallel} \
%B^2_{\infty}} \,=\,{3 \pi \over K\Phi }\xi_{\rm CR}{V_{\rm sh}\over c}{\rho
%V_{\rm sh}^2\over P_{B_\infty}}{1\over y k_\parallel},
%\end{equation}
%with $P_{B_{\infty}}=B_\infty^2/8\pi$ the magnetic pressure in the
%supernova environment, and $K\sim \log(k_{\rm max}/k_{\rm min}$ a
%number of order $10$. The saturation level $k_{\parallel}
%S(k_{\parallel})$ can be quite high (see discussion below), i.e. for
%extreme shock velocities $V_{\rm sh}/c \sim 0.1$, the above implies
%$k_\parallel S \approx$ at $y=1$.

However, the instability may also saturate through non-linear transfer
effects. As the field builds up through the instability, the
non-linear transfer time $t_{\rm n-lin}(k_\parallel)$ along the
$k_\parallel$ direction decreases to the point where the instability
saturates when $G_{\rm n-res}(k_\parallel)t_{\rm
n-lin}(k_\parallel)=1$. In order to see when this happens, one can
express the non-linear transfer time as:
\begin{equation}
t_{\rm n-lin}(k_\parallel) \,=\, \left[k_\parallel \overline{V}_{\rm
A}(k_\parallel)\right]^{-1}\ ,
\end{equation}
with $\overline{V}_{\rm A}(k_\parallel) = \delta
B(k_{\parallel})/\sqrt{4\pi n_0}$, and the non-resonant growth rate
$G_{\rm n-res}(k_\parallel)=\sqrt{k_\parallel k_{\rm
c}}\,\overline{V}_{\rm A}$. We then find for the saturated field at a
scale $k_{\parallel}$:
\begin{equation}
\label{Eq:delB}
\delta B^2(k_\parallel)\simeq {12\pi\over\Phi}P_{\rm CR}{V_{\rm
sh}\over c}{1 \over k_\parallel r_\star} \ .
\end{equation}
Integrating this result over $k_{\parallel}r_* >1$, we exactly obtain
the same saturation level as previously, which is quite
remarkable. This means that even if there is a kind of quasi-linear
saturation at work, within the same time the non-linear transfer
remodels the spectrum. The spectrum derived above from this remodeling
process is likely correct but would require a more elaborated theory
together with sophisticated numerical simulations to be confirmed.

Using Eq.(\ref{Eq:delB}) the non-resonant spectrum profile is thus
\begin{equation}
\label{Eq:Satnr}
S(k_\parallel,x) = \ \frac{2\pi \ \delta B^2(k_\parallel)}{k_\parallel
\ B^2_{\infty}} = {3\pi\over\Phi} \xi_{\rm CR} {V_{\rm sh}\over c}
{\rho_0 V_{\rm sh}^2 \over P_{B_\infty}} {1 \over r_\star k_\parallel^2} \propto
k_{\parallel}^{-2} \ .
\end{equation}

We obtained this result by assuming energy transfer along the
$k_\parallel$ direction. However the relevant transfer could be
transverse through an Alfv\'en cascade, as is often considered. In
this case the transfer rate $k_{\perp} \bar u$ is supposed to be
faster and a balance $k_{\perp} \bar u \sim G_{\rm
n-res}(k_{\parallel})$ is expected. However it generally requires the
prescription of a relation between $k_{\perp}$ and $k_{\parallel}$
[see subsection(\ref{S:alfven_turb})].  Setting $k_{\perp}r_{\star}
\sim (k_{\parallel}r_{\star})^m$ with $m\geq 1$, we obtain a spectrum
\begin{equation}
\label{Eq:Strans}
S(k_{\parallel}) \propto k_c r_{\star}^2(k_{\parallel}r_{\star})^{-2m} \ .
\end{equation}
There is no clear constraint on the value of $m$, which may be taken
to $1$ without apparent inconsistency. In any case, as $m \geq 1$, the
same estimate (\ref{Eq:Bsatnr}) of the saturation level is found,
which makes this estimate fairly robust.

To draw a complete picture of the saturation spectrum, one would need
to follow the evolution of non-linear turbulence transfer at the same
time as the evolution of the non-resonant instability and the position
dependence of $r_\star$.

\subsubsection{The resonant regime}
\label{S:satr}
As the turbulence is advected to the shock front, cosmic rays with
Larmor radius $r_{\rm L}=1/k_\parallel$ appear and induce the resonant
instability. This latter is then quenched by advection through the
shock, which thus provides the main saturation mechanism of this
instability.

i) For the sake of clarity, we first assume equally amplified forward
and backward spectra, and $S=S^++S^-$. The evolution equation for this
latter reads
\begin{equation}
\label{Eq:comp}
{\partial\over\partial x}S(k_\parallel,x)\,=\,-a_{\rm
res}(k_\parallel,x){e^{-x/\ell_{\rm d}}\over \ell_{\rm d}}S(k_\parallel,x)\ .
\label{eq:res}
\end{equation}
The initial condition for integration lies at $x=x_{\rm
min}(k_\parallel)$ and is given by the spectrum produced by the
non-resonant instability. We argue that the resonant amplification
will not modify the overall magnetic field strength $\overline B$ over
that produced the non-resonant instability by a large factor for
typical SNe environment and shock wave values. Hence, to solve
Eq.~(\ref{Eq:comp}) above, we first assume that the ratio
$(1-\eta)^{-1}$ is constant between the shock front ($x=0$) and the
point where the resonant instability first comes into play, defined by
$x=x_{\rm min}(k_\parallel)$ for $k_\parallel=k_{\rm min}$. Then $1/
(1-\eta)^{1/2} =\overline B_*/ B_\infty$ with $\overline B_*$ the
value of $\overline B$ at this latter point. In this case, $\ell_{\rm
D}$ does not depend on $x$ in the interval in which the resonant
instability acts, and the equation for $S$ can be solved as:
\begin{equation}
S(k_\parallel,x)\,=\,S(k_\parallel,x_{\rm min}) + {\tilde{\alpha}_{\rm
res} \over k_{\parallel}} \left(e^{-x/ \ell_{\rm
D}}-e^{-1}\right){\overline B_*\over B_\infty}\ ,
\end{equation}
with $\tilde{\alpha}_{\rm res}\equiv (\pi / \rm \Phi){\cal
M}_{A\infty}\xi_{\rm CR}$ (see Eq.(\ref{Eq:a_res})).  The term $S_{\rm
nr}\equiv S(k_\parallel,x_{\rm min})$ is the spectrum produced by the
non-resonant instability and is proportional to $k_\parallel^{-2}$ due
to saturation. The second term on the rhs is the contribution of the
resonant instability and we denote it $S_{\rm r}(k_\parallel)$ at the
shock front; it is proportional to $k_\parallel^{-1}$ and dominates
the former over the whole range of wavenumbers if $a_{\rm
res}\overline B_*/B_\infty$ is sufficiently large.  In case of the
absence of non-resonant instability, $\overline{B}_* = B_{\infty}$ and
the expression of $\tilde{\alpha}_{\rm res}$ leads to a resonant
saturation level (see \citet{BL})
\begin{equation}
\label{Eq:Bres}
{B_{\rm res}^2\over B^2_{\infty}}\,\sim\, \xi_{\rm CR} M_{\rm A\infty} \ .
\end{equation}
The contribution of the resonant instability to the magnetic field
energy density $B_{\rm res}^2/8\pi$ is obtained from:
\begin{equation}
\int_{k_{\rm min}}^{k_{\rm max}}{\rm d}k_{\parallel}\, S_{\rm
r}(k_\parallel)\,\sim\, \tilde{\alpha}_{\rm res}{\overline B_*\over
B_\infty}\log\left({k_{\rm max}\over k_{\rm min}}\right)\ .
\end{equation}
Hence
\begin{equation}
\label{Eq:rapb}
{B_{\rm res}^2\over B_{\rm n-res}^2}\ = {\pi \over \Phi}
\log\left({k_{\rm max}\over k_{\rm min}}\right)\ \xi_{\rm CR} M_{\rm
A\infty} {B_{\infty} \over \overline{B}_*} \sim\, \sqrt{{\xi_{\rm CR}
\ c \over V_{\rm sh}}}\ ,
\end{equation}
this ratio is larger than one by a factor of a few for shock
velocities lower than a few times $\sim \xi_{\rm CR} c$.  Among other
things, this implies that the scaling with $k_\parallel$ of the total
spectrum at the shock front will be dominated by the resonant
contribution.
%\begin{equation}
%{B_{\rm res}^2\over B_{\rm n-res}^2}\,\sim\,{\overline V_{\rm A}c\over 
%V_{\rm sh}^2}\log\left({k_{\rm max}\over k_{\rm min}}\right)\ .
%\end{equation}

One can investigate the effect of the above assumption
$1/(1-\eta)=\,$constant by integrating formally Eq.~(\ref{eq:res}) as:
\begin{equation}
S(k_\parallel,x) \,=\, S(k_\parallel,x_*) + {a_{\rm res}\over
k_\parallel} \int^{1}_0 {\rm d}y\, e^{-y}{\overline
B[x(y,k_\parallel)]\over B_\infty}\ ,
\label{eq:res2}
\end{equation}
with $y = x/\ell_{\rm d}(x)$.  The integration in the rhs can be
understood as a function of $k_\parallel$ so that the dependence of
the second term on the rhs is not strictly speaking $\propto
k_\parallel^{-1}$. However $\overline B(x)/B_\infty$ is a decreasing
function of $x$ and therefore it is bounded by below by $\overline
B[x_*(k_\parallel=k_{\rm min})]/B_\infty$ and by above by $\overline
B(0)/B_\infty$. The integral in Eq.~(\ref{eq:res2}) is thus bounded by
two constants that are independent of $k_\parallel$ and whose ratio is
a factor $\overline B(0)/\overline B[x_*(k_\parallel=k_{\rm
min})]\sim\,$ a few. Hence the integral modulates only weakly the
powerlaw $\propto k_\parallel^{-1}$ and we conclude that, at the shock
front, the spectrum $S\propto k_{\parallel}^{-1}$. It should be noted
that the total variation of the magnetic field in the resonant region
is even more limited if $\varepsilon > 0$ as we shall see in paper II,
as in that case $a_{\rm res}$ varies as
$(\overline{B}/B_{\infty})^{1-\varepsilon}$.\\ ii) If the backward
spectrum is not amplified nor remodeled by mode coupling, the solution
is similar to the previous case without the factor 2.\\ iii) If the
backward waves are damped at the same rate as the forward waves are
amplified, $S^+ S^- = \rm{constant}$.  If the forward spectrum is
amplified an obvious solution is obtain for $S^+ \gg S^+(x=x_{\rm
min})$
\begin{eqnarray}
\label{SPSM}
S^+ & \simeq & S_*(k_\parallel) \times (\exp(-x/\ell_{\rm d})-e^{-1})
\nonumber \\ S^- & \simeq & \frac{S^+(x=x_{\rm min})S^-(x=x_{\rm
min})}{S^+}
\end{eqnarray}
The backward waves are damped exponentially. \\
iv) The backward waves can also be generated by backscattering process
of forward Alfv\'en waves off acoustic waves (slow magneto-sonic modes
precisely). This process deserves a specific development presented in
section 4.  

\subsection{The behavior of Alfv\'enic turbulence}
\label{S:alfven_turb}
The behavior of moderate MHD turbulence -- moderate in the sense that
a significant mean magnetic field is preserved -- is peculiar when
incompressibility is assumed, because of the particularity of Alfv\'en
waves dynamics. Resonant three wave interactions do not develop as
usual dispersive waves, because of their specific dispersion relation:
$\omega = k_{\parallel}V_{\rm A}$. However, a forward wave and an
opposite backward wave can couple through a resonant interaction with
a third wave with $k_{\parallel} = 0$ \citep{BN,GN}. The weak
turbulence description shows that the energy cascade in the inertial
range occurs only in the transverse direction to the mean field. The
stationary spectrum is in $k_{\perp}^{-3}$ and the dependence in
$k_{\parallel}$ is arbitrary, which means that it is determined by the
mechanism of generation of the turbulence. This behavior has been
observed in numerical simulations even in the regime of moderate
turbulence (\cite{BN}). The extension of the resonant three wave
interaction by taking account of a nonlinear broadening due to the
relaxation of triple correlation -- the so-called Eddy Damping Quasi
Normal Markovian description -- has been done for MHD turbulence with
a mean field by \citet{GS}. They have argued that some re-organization
of the spectrum occurs in $k_{\parallel}$ due to scaling constraints
between the nonlinear transfer in the transverse direction and the
parallel propagation of Alfv\'en waves.  Let us summarize this
discussion. These Alfv\'en waves (also called shear Alfv\'en waves, as
opposed to MHD waves that have parallel components) are incompressible
and purely transverse to the mean field. The turbulent energy density
$\epsilon\propto\bar u_{\perp}^2$ and the eddy turn over time
$\tau_{\rm n-lin} \sim (k_{\perp} \bar u_{\perp})^{-1}$. The scale
invariant spectrum of the energy cascade is unavoidably anisotropic
$S_{3\rm D}(k_{\perp}, k_{\parallel}) \propto k_{\perp}^{-q}
k_{\parallel}^{-\beta}$. The critical balance assumption of Goldreich
and Shridar is that the transfer rate $\tau_{\rm tr} = \tau_{\rm
n-lin} \sim \tau_{\rm A}$ at all scales, where $\tau_{\rm A} =
(k_{\parallel} V_{\rm A})^{-1}$. Then for an anisotropic inertial
cascade such that the energy transfer rate at each scale is constant,
namely $Q \propto \epsilon/\tau_{\rm n-lin}\propto k_{\perp}
u_{\perp}^3 = \, {\rm constant}$, a relation between parallel and
transverse wavenumbers is found:
\begin{equation}
\label{ }
k_{\parallel} \sim \frac{Q^{1/3}}{V_{\rm A}} k_{\perp}^{2/3} \ .
\end{equation}
In this more elaborated description (EDQNM) no energy transfer from
forward waves to backward waves and vice versa takes place. Recent
numerical simulations \citep{CV,MG} have suggested that the scaling
$\tau_{\rm n-lin} \propto \tau_{\rm A}$ was preserved in all regimes,
so that $\tau_{\rm n-lin} =\tau_{\rm A}/\chi$, with $\chi$ a constant
independent of the wavenumber at all scale, and $\tau_{\rm tr} =
\tau_{\rm n-lin}/\chi$. The previous relation is thus extended to
\begin{equation}
\label{KK}
k_{\parallel} \sim \frac{Q^{1/3}}{\chi^{4/3}V_{\rm A}} k_{\perp}^{2/3}
\ .
\end{equation}
Whereas the weak turbulence theory leads to a spectrum in
$k_{\perp}^{-2} f(k_{\parallel})$ with an arbitrary function $f$ of
$k_{\parallel}$, the Goldreich -Shridar theory leads to a spectrum
$S_{3\rm D}\propto k_{\perp}^{-q-2/3}
f(k_{\parallel}/k_{\perp}^{2/3})$. When a scale invariance in
$k_{\parallel}$ is generated in the turbulence situation, the spectrum
is of the form $S_{3\rm D} \propto k_{\perp}^{-q}
k_{\parallel}^{-\beta}$. Then Eq.~(\ref{KK}) together with the
assumption of a constant energy transfer rate $Q\propto
\epsilon/\tau_{\rm n-lin}$ with $\epsilon\sim k_\perp^2k_\parallel
S_{\rm 3D}$ provide a relation between the index of the parallel
spectrum with the index $\alpha$ of the perpendicular spectrum
\citep{GPM}:
\begin{equation}
\label{ }
3\alpha+2\beta = 7 \, \, {\rm with} \, \alpha = q-1 \ .
\end{equation}
This is considered to be the generalization of Iroshnikov-Kraichnan
theory \citep{IR,KR} when anisotropic effects are taken into
account. Still some arbitrariness is maintained. However the
CR-instability in resonant regime generates a turbulent spectrum such
that $\beta = 1$, and the transverse Alfv\'enic couplings between
modes then lead to $\alpha = 5/3$. Only couplings with slow
magneto-sonic modes may allow to obtain the same spectrum for the
backward waves and the slow waves.
\section{Nonlinear generation of backward waves}
\label{S:nlback}
The process $A^+ \rightarrow A^- + S^+$, where $A$ represents
Alfv\'enic modes and $S$ a slow magneto-sonic mode, is the only
process that can transfer energy from forward waves to backward waves
and it turns out to be efficient, as will be seen further on. The
frequency of the slow magneto-sonic mode is such that $\omega^s = k^s
V_{\rm sm}(\theta_s)$; for convenience, we write it $\omega^s =
\beta_s k^s_{\parallel}V_{\rm A}$, where, for $c_{\rm s} < V_{\rm A}$,
$\beta_s \simeq {c_{\rm s} \over V_{\rm A}}(1+{c_{\rm s}^2 \over
V_{\rm A}^2} \sin \theta_s)^{-1/2}$; this number is assumed smaller
than unity and weakly varying with $\theta_s$. The process is most
efficient under the resonance condition : $\omega^+ -\omega^-
-\omega^s = 0$. Since the wave vectors are such that $\vec k^+ -\vec
k^- -\vec k^s = 0$, we obtain the following relations between the
parallel wavenumbers:
\begin{eqnarray}
k_{\parallel}^+ & = & k_{\parallel}^s \frac{1+\beta_s}{2} \\
k_{\parallel}^- & = & - k_{\parallel}^s \frac{1-\beta_s}{2} < 0 \ .
\end{eqnarray}
Therefore, when the magnetic field is above the equipartition value
(e.g. $V_{\rm A} > c_{\rm s}$), $\beta_s < 1$, and we always get a
backscattering of Alfv\'en waves off slow magneto-sonic
modes. Backscattering would not be possible with other MHD waves, for
obvious kinematic reasons. This backscattering process with Alfv\'en
waves is analogous to the Brillouin backscattering process with usual
electro-magnetic waves of the vacuum. Even if no sonic waves are
excited beforehand, the primary Alfv\'en waves can generate them
spontaneously above some threshold [see \citet{PK}]. In the
interstellar medium $V_{\rm A} \simeq 3 \, c_{\rm s}$; it is already
sufficient to get the backscattering process.  The domination of the
Alfv\'en velocity over the sound speed is even increased at the
external shock of SNr because of their convexity.  Indeed in the shock
frame, the ambient medium converges towards the front at a velocity
$-\vec V_{\rm sh}$ that points towards the center of curvature, the
density increases and therefore the frozen in magnetic field has an
amplified transverse component.  From the evolution equations
multiplied by $\tau_{\rm adv}$, we introduce the dimensionless
parameter
\begin{equation}
\label{Eq:KAP}
\kappa \equiv \frac{\pi}{12}\frac{c\overline V_{\rm A}}{\beta_s V_{\rm sh}^2} \ ,
\end{equation}
which measures the importance of the backscattering process as
compared to advection. Typically, $\kappa$ is a number close to one.\\
For the most interesting cases where the backscattering process
efficiently remodels the stable spectra, the asymptotic spectra,
determined externally by the turbulence in the interstellar medium,
can be ignored. The spectra are then proportional to
$S_*(k_{\parallel})k_{\perp}^{-q}$. Because the diffusion length is
generally not constant, but dominated by the spectrum of unstable
waves, for numerical simulation purpose, it is convenient to describe
the profiles of the wave spectra with the help of a dimensionless
variable $y$ defined by $dx = \ell_{\rm D}(r_{\rm L}=1/k_{\parallel}, x)dy$.
Then the function $\phi = e^{-y}$. We have to bear in mind that, when
the problem is solved for the variable $y$, we can reconstruct the
spectrum profile in the variable $x$. Since $\ell_{\rm D}(r_{\rm
L}=1/k_\parallel,x) = (1/3)(c/V_{\rm sh})k_\parallel^{-1}
(k_\parallel/k_{\rm min})^{\beta-1}\eta^{-1}(x)$, one finds:
$$y = 3{V_{\rm sh}\over c}\left({k_\parallel\over
k_{\rm min}}\right)^{2-\beta}\int_0^{x}{\rm d}x' k_{\rm min} \  \eta(x'), 
$$
The quantity $r_*(x)$ is defined by $y=1$ for $r_{\rm L}=r_*$, hence
$y=\left(k_\parallel r_*\right)^{2-\beta}$. The regions with $y \ge 1$
($y \le 1$) correspond to far (close) distances to the shock front and
is dominated by the non-resonant (resonant) waves.  Therefore the
evolution of the spectra reduces to a differential system that governs
the evolution of their amplitude as a function of the $y$ variable. As
long as $\kappa$ is small, the solution given by Eq.(\ref{SPSM}) is
slightly modified and the order three system that describes the
generation of backward A-wave and forward S-wave is sufficient (see
appendix A). The numerical integration leads to the solutions sketched
on figure \ref{Fig1}, that shows that for, increasing $\kappa$, more
and more conversion into backward A-waves and forward S-waves is
realized. However when $\kappa$ is increased significantly, one has to
take account of the secondary process where backward A-waves decay
into forward A-waves and backward S-waves. The evolution is then
described by a system of order four (see appendix B). The numerical
solutions are displayed on figure \ref{Fig2}. It can be that a
significant backward spectrum is generated; however without changing
the order of magnitude of the primary spectrum.

\begin{figure}
\begin{tabular}{cc}
  \includegraphics[width=0.5\textwidth]{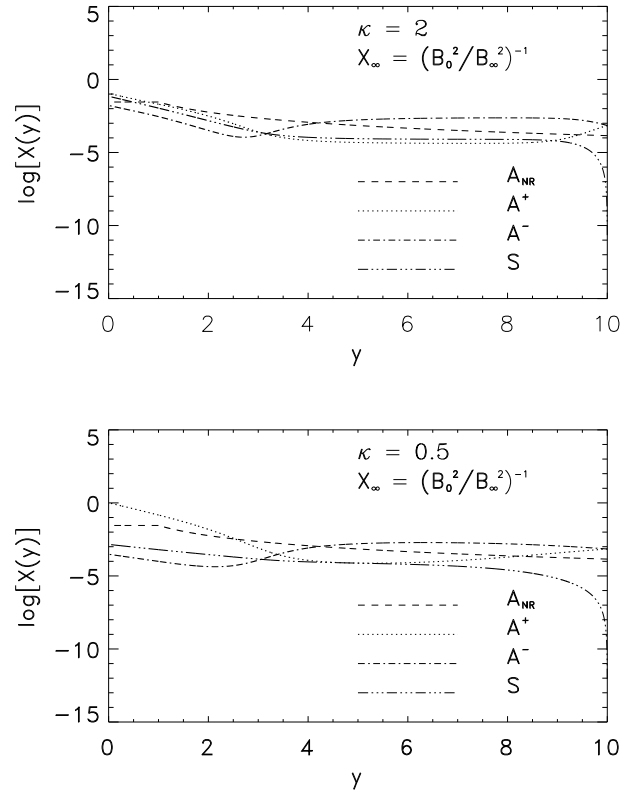} \end{tabular}
 \caption{Solutions of third order system for different value of the
 $\kappa$ parameter (see the definition in the text). Even if they have
 been defined for $y$ between 0 and 1, the resonant waves profiles are
 calculated between the shock front and $y = 10$ where they match
 their asymptotic interstellar values. The wave-particle resonance depends 
 on the particle pitch-angle $\alpha$, at $r_{\rm L} = r_*$ we have 
 $k_{\parallel} r_* \cos\alpha \simeq 1$, the product
 $k_{\parallel} r_*$ and then $y$ can be above 1.  In the upper panel,
 $\kappa =1$ is high leading to a strong conversion of forward
 Alfv\'en waves into backward Alfv\'en waves and sound
 waves. For $ y \le 5$, the resonant instability takes over the
 non-linear transfer, the forward Alfv\'en waves are produced and
 the backward waves are pumped. The sound waves are heavily produced
 between $y = 5$ and the shock front. The ratio of forward to backward
 Alfv\'en waves at the shock front is about three orders of
 magnitude. In the lower panel $\kappa =0.1$, the production
 of backward Alfv\'en and sound waves is less intense. 
 In both cases the amplification factor $\overline{B}_*/B_{\infty} = 10.$
 The boundary conditions are: $X^+ = X^- = X_{\infty} = A^{-2} \ll 1 $ and $X^s =
 0$. All simulations have been performed with $\epsilon = 0$.}
\label{Fig1}
\end{figure}

\begin{figure}
\begin{tabular}{cc}
  \includegraphics[width=0.5\textwidth]{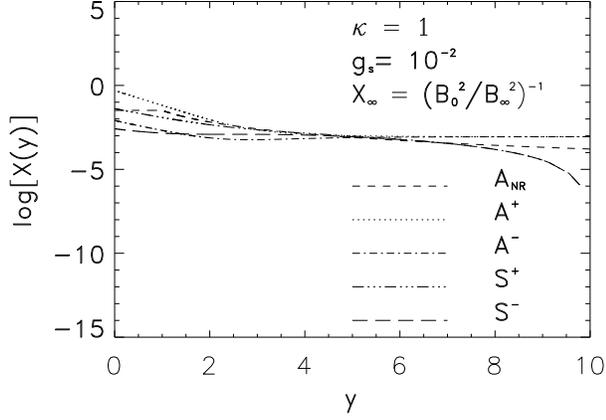} 
 \end{tabular}
\caption{Solutions of the fourth order system. The non-linear transfer
parameter $\kappa = 1$ and the damping rate of sound waves is $g_s = 5 \times
10^{-3}$. This choice corresponds to a ratio $V_{\rm A}/c_{\rm s} =
3$. All simulations have been performed with $\epsilon = 0$.  The results show
that as long as the damping or growing effects do not dominate the
transfer among the waves (for $y \le 5$), the non-linear transfer
mostly produce backward and forward sound waves equally. 
In both cases the amplification factor $\overline{B}_*/B_{\infty} = 10$. The boundary
conditions are: $X^+ = X^- = X_{\infty}=  A^{-2} \ll 1$, and $X^{s+} = X^{s-} =
0$.}
\label{Fig2}
\end{figure}

\section{Downstream: dynamo action and turbulence relaxation}
If turbulence is still moderate downstream, then the spectra built 
upstream are transmitted across the front, and thus a
$k_{\parallel}^{-1}$ 1D-spectrum is maintained downstream. Bohm
diffusion would then applies downstream as well.

\subsection{Helicity and estimate of dynamo amplification downstream}

The non-resonant regime of the streaming instability induces a
left-right symmetry breaking. Therefore the turbulence carries
helicity which offers grounds for dynamo action. The helicity can be
calculated in term of the difference between the spectrum of
right-handed modes $S_{\rm RH}$ and the spectrum of left-handed modes
$S_{\rm LH}$ (the $k_{\perp}$ dependence has been integrated out):
\begin{equation}
\label{ }
H \equiv \langle\vec u \cdot rot\, \vec u\rangle = 2V_{\rm A}^2 \int (S_{\rm
RH}-S_{\rm LH})k_{\parallel} \frac{{\rm d}k_{\parallel}}{2\pi} \ .
\end{equation}
The integrand can be considered as the helicity spectrum $S_H$. This
spectrum is used to calculate the so-called "alpha"-parameter of the
turbulent dynamo theory:
\begin{equation}
\label{Eq:alphad1}
\alpha_{\rm D} = \int
\frac{\Gamma(k_{\parallel})}{\omega_k^2+\Gamma^2(k_{\parallel})}
S_H(k_{\parallel})\frac{{\rm d}k_{\parallel}}{2\pi} \ ,
\end{equation}
where $\Gamma$ is the damping rate of the turbulence in stationary
state. In our problem the main damping mechanism is the shock
advection: $\Gamma(k_{\parallel}) = 1/\tau_{\rm
adv}(k_{\parallel})$. For the non-resonant modes $\omega_k^2 \ll
\Gamma^2(k_{\parallel})$ and the dynamo coefficient reads:
\begin{equation}
\label{Eq:alphad2}
\alpha_{\rm D} = \frac{2}{3} c\frac{V_{\rm A}^2}{V_{\rm sh}^2} \int
\frac{S_{\rm RH}-S_{\rm LH}}{S^++S^-}
\frac{1}{k_{\parallel}}\frac{{\rm d}k_{\parallel}}{2\pi} =
\frac{2c}{3\pi} \frac{V_{\rm A}^2}{V_{\rm sh}^2} \ln \frac{r_*}{r_0} \ ,
\end{equation}

The helicity is transfered through the shock as has been calculated by
\citet{SC}. Helicity in the spectrum matrix leads to a third
diffusion coefficient for the cosmic rays because the two transverse
space variable are correlated $\langle\Delta x_1 \Delta x_2\rangle \neq 0$ (see
paper II). 

The mean field evolves in the turbulent plasma according to
the following equation:
\begin{equation}
\label{ }
\frac{\partial}{\partial t} \vec A = \alpha_{\rm D} \vec B + \vec u \times
\vec B + \nu_t \Delta \vec A \ ,
\end{equation}
where $\nu_t$ is the turbulent magnetic diffusivity. A typical scale
for the variation of the mean field arises, namely $\ell_{\rm dyn} =
\nu_t/\alpha_{\rm D}$ with an associated time scale $\tau_{\rm dyn} =
\ell_{\rm dyn}/u_2$. More precisely, the dynamo modes of wavelength
larger than $\nu_t/\alpha_{\rm D}$ grow and it is expected that the
mean field reaches an intensity on the order of the equipartition
value, not more.

\subsection{Relaxation or compression downstream}
The turbulence properties downstream (level of turbulence, spectral
index) can be constrained from the size of the X-ray filaments in
young SNr [see \cite{Petal05}]. It is shown that the relativistic
electrons with tens of TeV energies producing the observed synchrotron
radiation have a diffusion coefficient close to the Bohm
value. Recently, \citet{Pohletal05} have pointed out the importance of
the relaxation length of the turbulence downstream the shock. The
authors stressed that the size of the X-ray filaments observed in
young SNr (see the discussion in section \ref{S:astro}) may be well
controled by the turbulence rather than by the synchrotron
losses. However, the previous analysis has been made assuming an
isotropic turbulence spectrum, which is not correct at least for two
reasons: the turbulence is already anisotropic upstream and the
magnetic field amplified upstream is compressed in the direction
parallel to the shock front. 

In order to elucidate the way the turbulence acts on relativistic
particles, we compare the non-linear Alfv\'en transfer time $t_{\rm
n-lin}(k_\parallel) = [k_\parallel \overline V_{\rm
A}(k_\parallel)]^{-1}$
\begin{equation}
\label{Eq:tnl}
t_{\rm n-lin}(k_\parallel) \,=\, {l_\parallel \over
\overline{V}_{\rm A} \sqrt{\beta-1}}(l_\parallel
k_\parallel)^{(\beta-3)/2} \ ,
\end{equation}
and the term $\beta-1$ in the prefactor should be replaced by
$1/\log(k_{\rm max}/k_{\rm min})$ when $\beta=1$.  The downstream
return timescale reads:
\begin{equation}
\label{Eq:tret}
t_{\rm ret}\,=\,\kappa \left({c \over V_s}\right)^2 
{2 \over 3}t_{\rm s},
\end{equation}
$t_{\rm s} = (2\pi)^{-\beta}\eta^{-1}(l_\parallel/c)(r_{\rm
L}/l_\parallel)^{2-\beta}$ being the angular scattering timescale
downstream; in the above expressions, $l_\parallel$ is the coherence
length in the parallel direction.  The prefactor $\kappa<1$ accounts
for the shortening of the return timescale in compressed
turbulence. The condition $t_{\rm ret} < t_{\rm n-lin}$ means the
particle does explore distances smaller than the relaxation length of
the turbulence downstream and experience a compressed rather than a
relaxed turbulence during their journey downstream. In the case of
Bohm type turbulence (meaning $\beta=1$) the ratio of the two
timescales translates into
\begin{equation}
\label{Eq:treltnl}
\frac{t_{\rm ret}}{t_{\rm n-lin}} \ \simeq \ \frac{\kappa}{3\pi} \
\frac{c}{V_{\rm sh}{\cal{M}}_{\rm A\infty}} \frac{\log(k_{\rm
max}/k_{\rm min})^{-1/2}}{(1-\eta)^{1/2} \eta} \ ,
\end{equation}

%\begin{equation}
%\label{Eq:treltnl}
%\frac{t_{\rm ret}}{t_{\rm n-lin}} \ \simeq \ \frac{2\kappa}{3} \
%\frac{c}{V_{\rm sh}} \frac{1}{(1-\eta)^{1/2} \eta \ \Phi^{1/2}
%{\cal{M}}_{\rm A}} \ .
%\end{equation}

Using typical values of the magnetic field, mean density and shock
velocity in our problem (see next section) and acknowledging for a
saturation level of ${\cal{M}}_{\rm A\infty} \xi_{\rm CR}
\overline{B}_*/B_{\infty} \simeq \eta/(1-\eta) \simeq 1/(1-\eta)$, we
find a ratio $t_{\rm ret}/t_{\rm n-lin} \simeq 0.5 \kappa \
(V^{-1}_{\rm sh,-1} \xi_{\rm CR} / \Phi)^{3/4}$ where $V_{\rm sh,-1}$
is the shock velocity in units of $0.1c$. The previous ratio is thus
expected to be $\le 1$ unless the shock velocity is lower than
$10^{-2}c$ .

\section{Astrophysical consequences}
\label{S:astro}
Supernova blast waves explore different external interstellar media
(ISM) during their evolution. We consider only two phases of the SN
evolution: the very early free expansion phase where the shock
velocity can reach extreme values as high as $V_{\rm sh} \simeq 0.1 c$
and the late free expansion phase (or early Sedov self-similar phase)
where the shock velocity drops to $V_{\rm s} \simeq 10^{-2} c$. These
two phases are the more relevant concerning high energy cosmic ray
production. In the latter phase the remnant may expand either in a hot
rarefied ($\rm{T} \simeq 10^6 \ \rm{K}, \ \rm{n} \simeq 10^{-3/-2} \
cm^{-3}$) interior of a massive star wind bubble or in a warm
partially ionised ($\rm{T} \simeq 10^4 \ \rm{K}, \ \rm{n} \simeq
10^{-1/0} \ cm^{-3}$).  The mean ISM magnetic field in both cases is
conservatively taken to $3\mu \rm{Gauss}$. In the very early phase of
the SNr evolution, the medium is probably much denser with $\rm{n}
\simeq 10-100 \ cm^{-3}$.

 The ratio of the saturation magnetic field energy density of the
non-resonant and resonant instability (when both regimes are present)
respectively is $S_{\rm n-res}/S_{\rm res} \simeq \sqrt{V_{\rm sh}/
(\xi_{\rm CR} \ c)}$.  The non-resonant regime appears to dominate for
the very early free-expansion phase as already pointed out by
\citet{AB}, while the resonant regime dominates by a factor $10-100$
(see discussion in section \ref{S:satr}) in the late free-expansion
phase and Sedov self-similar phases.  The magnetic field deduced (only
lower limits) from the size of bright X-ray filaments in young SNr
\citep{BER,V04,Vetal05,Petal05} is expected to be mostly produced in
the resonant regime should then scale approximately as $\sqrt{V_{\rm
sh}}$ [see Eq.(\ref{Eq:Bres})]. We have seen above that if the
non-resonant regime contributes substantially to the amplification
this dependence is not as simple and the way the shock decelerates
during the earlier phases can modify it.  The amplification by the
non-resonant instability may lead in the most extreme cases to very
high amplification levels, pointing towards SNr in very early
free-expansion phase as efficient CR accelerators.  In that case, the
magnetic density should scale as $V_{\rm sh}^3$ as pointed out by
\citet{AB}. This issue is of prime importance and should deserve
detailed observational investigations, unfortunately difficult to
perform in this SNr evolution stage. However, this early phase lasts
for a very small fraction of the whole SNr lifetime except in a low
density and highly magnetised medium as expected in a turbulent hot
ISM phase often called as superbubbles [see for instance
\citet{Petal04}]. Answering the question of the maximum CR energy
expected in SNr and the origin of the CR knee at $\sim 3 \times
10^{15}\,$eV requires then a time dependent CR spectrum calculation
[see \citet{PZ05}] and to account for the CR diffusion regimes
correctly.  If the first point is beyond the scope of the present
work, the second point will be discussed in paper II.

\section{Conclusion}
We summarize the main results of the work as follows. Upstream of an
astrophysical shock, the cosmic ray streaming triggers an instability
that has two different regimes: one occurs under resonant condition
and dominates the longer wavelengths of the Alfv\'en spectrum, the
other occurs off resonance and dominates at shorter wavelengths [this
is the Bell regime of the instability \citep{AB}]. In the purpose of
investigating the turbulent transport of the highest energy cosmic
rays both regimes have to be considered over the remnant
evolution. The non-resonant instability saturates either by non-linear
transfer effects or by a quenching effect at $k_{\rm c} r_* = 1$. The
saturation level is $k_{\parallel} S(k_{\parallel}) \simeq \xi_{\rm
CR}(V_{\rm sh}/c) \ {\cal{M}}^2_{\rm A\infty} / (\Phi k_{\parallel}
r_\star)$. The main saturation mechanism for the resonant instability
stems from the fact that the shock front catches up with the growing
waves over a diffusion length. The saturation level is modulated by
the non-resonant saturation level according to: $k_{\parallel}
S(k_{\parallel}) \simeq \xi_{\rm CR} \ ({\cal{M}}_{\rm A\infty} / \Phi)
\times \overline{B}_{\rm n-res}/B_{\rm \infty}$.  The non-resonant
regime of the instability dominates the resonant contribution only for
very fast shock velocity.

%compared to $\sqrt{\overline V_A c}$.

The streaming instability partially determines the spectrum, namely
its $k_{\parallel}$ dependence. The spectra are close to
$k_{\parallel}^{-1}$, which, as will be shown in the second paper, can
lead to a Bohm scaling for the transport of cosmic rays.  The
$k_{\perp}$ dependence of the spectrum is remodeled by the non-linear
cascade of Alfv\'en waves, that essentially works transversally, the
transfer time being short enough as compared to the advection
time. The excited Alfv\'en turbulence constitutes the scattering
medium for the cosmic rays and it would be incomplete if only the
forward Alfv\'en waves would be present as a result of the resonant
streaming instability. A second nonlinear transfer develops which is
the backscattering of primary Alfv\'en waves off slow magneto-sonic
modes, and which re-distributes the energy from the forward Alfv\'en
waves to the backward ones and to the magneto-sonic ones. This process
turns out to be unavoidable because the Alfv\'en speed exceed the
sound speed upstream and is sufficiently fast compared to the
advection time.\\ Some turbulent dynamo action can be expected
downstream, but the intensity of the mean field should not
significantly exceed the equipartition value. The turbulence is also
compressed at the shock front producing reduced residence time of the
relativistic particles downstream. Apart for low shock velocities,
i.e. lower than $10^{-2}$ c, the residence time downstream is lower
than the non-linear transfer time controled by the Alfvenic
cascade. The turbulence spectra downstream are likely similar to those
that have been formed upstream, since they correspond to stationary
solutions of the turbulence equations in a mean field, according to
recent developments of the theory of Alfv\'enic turbulence. This
implies that the Bohm regime of cosmic ray transport if true upstream
would also apply downstream. Regarding the magnetic turbulence in the
Galaxy where a mean field imposes its constraint, a question rises
about the determination of the two indices $\alpha$ and $\beta$ of the
anisotropic spectrum.  They are only linked by the relation $3\alpha +
2\beta = 7$ \citep{GPM}. It is reasonable to think that $\beta$ and
thus $\alpha$ are determined by shocks and thus $\beta = 1$ and
$\alpha = 5/3$ (anisotropic Kolmogorov spectrum) would be
ubiquitous. \\

\section{Acknowledgment}

One of us, G.P., acknowledges fruitful discussions with Andrew Bell
and Heinrich V\"olk. We are also grateful to S\'ebastien Galtier for
exchanges about the most recent understanding of Alfv\'enic
turbulence.
\appendix
\section{The nonlinear operator of backscattering}
The energy spectra are normalized such that $\omega^a N^a =
S_{3\rm d}^a /\rho_0 V_{\rm A}^2$; in other words, the $N^a$ have the
dimension of an action (namely, an occupation number times $\hbar$).
\begin{eqnarray}
\dot N^+ &=&-\int \frac{{\rm d}^3 \vec k^-}{(2\pi)^3} \frac{{\rm d}^3
\vec k^s}{(2\pi)^3}w \nonumber\\
&&\times (N^+N^-+N^+N^s-N^-N^s) \\ 
\dot N^- &=&\int
\frac{{\rm d}^3 \vec k^+}{(2\pi)^3} \frac{{\rm d}^3 \vec
k^s}{(2\pi)^3} w  \nonumber\\
&&\times(N^+N^-+N^+N^s-N^-N^s) \\ 
\dot N^s &=&\int \frac{{\rm
d}^3 \vec k^+}{(2\pi)^3} \frac{{\rm d}^3 \vec k^-}{(2\pi)^3} w
 \nonumber\\
&&\times(N^+N^-+N^+N^s-N^-N^s)\ .
\end{eqnarray}
The transition probability has been calculated by \citet{AA} and reads
\begin{eqnarray}
\label{Eq:w}
w & = & \frac{\pi}{8} f \frac{V_{\rm A}^4}{\rho_0}
\frac{(k_{\parallel}^+)^2(k_{\parallel}^-)^2(k_{\parallel}^s)^2}
{\omega_+\omega_f\omega_s} \times  \nonumber\\
& & (2\pi)^3\delta(\vec k^+ -\vec k^-
-\vec k^s) \delta(\omega^+ -\omega^- -\omega^s)
\end{eqnarray}
with the angular factor $f$ depends on the unitary vectors $\vec n
\equiv \vec k/k$ and defined by
\begin{equation}
\label{Eq:f3n}
f = f(\vec n_+, \vec n_-, \vec n_s) \equiv \frac{((\vec n_+ \times
\vec e_b)\cdot (\vec n_- \times \vec e_b))^2}{(n_+^{\perp})^2
(n_-^{\perp})^2}
\end{equation}
where $\vec e_b$ is the unitary vector in the direction of the mean
field $\vec B_o$.  The transition probability $w$ can be rewritten in
the following way:
\begin{eqnarray*}
\label{Eq:w2}
w & = & {\pi \over 8} f \frac{\vert \cos \theta_- \vert}{\beta_s
   \rho_0} k_{\parallel}^+ \vert k_{\parallel}^-\vert k_{\parallel}^s
   \, (2\pi)^2 \delta(\vec k_{\perp}^+ - \vec k_{\perp}^- -\vec
   k_{\perp}^s) \times \mbox{}  \nonumber\\
 & & 2\pi\delta(k_{\parallel}^+
   -k_{\parallel}^s \frac{1+\beta_s}{2}) \delta(k_{\parallel}^- +
   k_{\parallel}^s \frac{1-\beta_s}{2}) \ .
\end{eqnarray*}
The last product of the two $\delta$-functions can be written under
several convenient forms for the calculation of the various
integrals. The control parameter $\kappa$ (Eq~\ref{Eq:KAP}) rises
after multiplying the nonlinear operator kernel by the advection time
$\tau_a$ [Eq.~(\ref{Eq:TOA})]. Furthermore, assuming unmodified
transverse spectra of the form $k_{\perp}^{-q}$ with $q>2$, when we
writes the system for $S^+$, $S^-$ and $S^s$, in term of the variable
$y$, it can be realized that it is scale invariant. Assuming power law
solutions for the S's, the coefficients of the system are independent
of the wave vectors after integrating over the angles. Because of the
integration of the delta-functions over the k's, the system is reduced
to a differential system involving three 1D-spectra depending on a
single wavenumber $k_{\parallel}$, since $k_{\parallel}^- \simeq
-k_{\parallel}^+$ and $k_{\parallel}^s \simeq 2 k_{\parallel}^+$.
Before writing the differential system, we approximate the theory
specifically for the case $y<1$, where we have already seen that the
spatial variation of the rms magnetic field is smooth compared to
$e^{-y}$ and we fix $\overline B$ at its value $B_0$ at the shock
front and introduce the amplification factor $A \equiv
B_0/B_{\infty}$. We set $N = X(y)N_*$ for the three spectra, where we
introduce $N_*$ such that $S_*(k_{\parallel}) 2\pi (q-2)l^2
(k_{\perp}l)^{-q} = k_{\parallel}V_{\rm A
\infty}N_*(k_{\parallel})/\rho_0 V_{\rm A \infty}^2$.

 The evolution system accounts for the case of substantial
pre-amplification by the non-resonant instability. The third-order
evolution system reads as
\begin{eqnarray}
(X^++X^-)\frac{\partial X^+}{\partial y} & = & -e^{-y}X^+ +  A\kappa \times\nonumber\\
& & [X^+X^-+(X^+-X^-)X^s] \\
(X^++X^-)\frac{\partial X^-}{\partial y} & = & e^{-y}X^- - A\kappa \times \nonumber\\
& & [X^+X^-+(X^+-X^-)X^s] \\
(X^++X^-)\frac{\partial X^s}{\partial y} & = & -{A\kappa  \over 2}\nonumber\\
& & \times [X^+X^-+(X^+-X^-)X^s]
\end{eqnarray}
For $\kappa = 0$, the ratio of the first two equations leads to
$X^+X^- = {\rm constant} = X^+(y=1)X^-(y=1)$.

\section{A more complete nonlinear theory}
Because of the efficiency of the backscattering process when $\kappa
\sim1$, it is reasonable to envisage a secondary generation of
backward sound waves from backward Alfv\'en waves, which also
regenerates the forward Alfv\'en spectrum: $A^- \rightarrow A^+ +
S^-$.
\begin{eqnarray}
\dot N^- & = & -\int \frac{d^3 \vec k^+}{(2\pi)^3} \int \frac{d^3 \vec
k^s}{(2\pi)^3} w \nonumber\\
&&\times(N^-N^++N^-N^{s-}-N^+N^{s-})\\ 
\dot N^+ & = & \int\frac{d^3 \vec k^-}{(2\pi)^3} 
\int \frac{d^3 \vec k^s}{(2\pi)^3}w\nonumber\\
&&\times(N^-N^++N^-N^{s-}-N^+N^{s-})\\ 
\dot N^{s-} & = & \int \frac{d^3 \vec
k^+}{(2\pi)^3} \int \frac{d^3 \vec k^-}{(2\pi)^3} w\nonumber\\
&&(N^-N^++N^-N^{s-}-N^+N^{s-})\ .
\end{eqnarray}
We combine the primary and the secondary process, include damping of
the stable waves (actually the backward waves are damped by the cosmic
ray streaming at the same rate as the forward waves are
amplified). The damping rate of the sound waves is
$\sqrt{\frac{\pi}{8}\frac{m_e}{m_i}} k_{\parallel}c_{\rm s}$.  We
proceed as in the previous appendix to describe the nonlinear
evolution of the resonant instability, and found that the spectra
proportional to $S_*$ are still recovered. We then form the four order
differential system on the amplitudes of the spectra that governs the
y-profiles:
\begin{eqnarray}
(X^++X^-)\frac{\partial X^+}{\partial y} & = & -e^{-y}X^+ - A \kappa \times \nonumber\\
& & (X^--X^+)(X_s^++X_s^-)  \\
(X^++X^-)\frac{\partial X^-}{\partial y} & = & e^{-y}X^- + A \kappa \times \nonumber\\
& &(X^--X^+)(X_s^++X_s^-) \\
(X^++X^-)\frac{\partial X_s^+}{\partial y} & = & g_sX_s^+ - A {\kappa \over 2} \times \nonumber\\
& & (X^-X^++X^+X_s^{+}-X^-X_s^{+})  \\
(X^++X^-)\frac{\partial X_s^-}{\partial y} & = & g_sX_s^- -A {\kappa \over 2} \times\nonumber \\
& & (X^-X^++X^-X_s^{-}-X^+X_s^{-})
\end{eqnarray}
In the case $\varepsilon = 0$, $g_s$ is a pure number: $g_s = A
\frac{2}{3}\sqrt{\frac{\pi}{8}\frac{m_e}{m_i}}\frac{cc_{\rm s}}{V_{\rm
sh}^2}$ which has to be compared with $\kappa$; typically $g_s \sim
(10^{-3}-10^{-2}) \kappa$. It turns out that a relaxation of the sound
waves is possible only for $g_s^2 > \kappa^2(X^+-X^-)^2$, which
implies $X^+ \simeq X^-$.


\begin{thebibliography}{}
\bibitem[Akhiezer \& Akhiezer(1975)]{AA} Akhiezer A., Akhiezer I.A., Polovin RV, Sitenko AG,
Stepanov KN, 1975, Plasma Electrodynamics, vol. 2 (Nonlinear Theory
and Fluctuations), Pergamon Press, Oxford.
\bibitem[Bell(2004)]{AB} Bell A.R., 2004, MNRAS, 353, 550.
\bibitem[Bell \& Lucek(2001)]{BL}Bell A.R., Lucek S.G., 2001, MNRAS, 321, 433.
\bibitem[Berezhko \& V\"olk(2004)]{BER} Berezhko E.G. \& V\"olk H.J., 2004, Astron. \& Astrophys., 419, L27.
\bibitem[Berezhko \& Ellison(1999)]{BE99} Berezhko, E.G. \& Ellison, D.C., 1999, ApJ, 526, 385
\bibitem[Berezhko et al(1996)]{Betal96} Berezhko, E.G., Elshin, V.K. \& Ksenofontov, L.C., 1996, JETP, 82, 1
\bibitem[Bhattacharjee \& Ng(2001)]{BN} Bhattacharjee A., Ng C.S., 2001, ApJ, 548, 318.
\bibitem[Casse et al.(2002)]{Cassetal02} Casse, F., Lemoine, M. \& Pelletier, G., 2002, Phys. Rev.D, 65, 023002
\bibitem[Drury et al.(2001)]{Druryetal01} Drury, L.O'C. et al, 2001, SSR, 99, 329
\bibitem[Cho \& Vishniac(2000)]{CV} Cho J., Vishniac E.T., 2000, ApJ 539, 273.
\bibitem[Galtier et al.(2000)]{GN} Galtier S., Nazarenko S.V., Newell A.C., Pouquet A.,2000, J. Plasma Phys., 63, 447.
\bibitem[Galtier et al.(2005)]{GPM} Galtier S., Pouquet A., Mangeney A., 2005, Physics of Plasma, 12, 092310. 
\bibitem[Goldreich \& Sridhar(1995)]{GS}Goldreich P., Shridar S., 1995, ApJ 438, 763.
\bibitem[Goldreich \& Sridhar(1997)]{GS2}Goldreich P., Shridar S., 1997, ApJ 485, 680.
\bibitem[Iroshnikov(1964)]{IR}Iroshnikov P.S., 1964, Soviet Astronomy 7, 566.
\bibitem[Kraichnan(1965)]{KR}Kraichnan, R.H., 1965, Phys. Fluids, 8, 1385
\bibitem[Lerche(1967)]{LER}Lerche I., 1967, Astrophys. J., 147, 689.
\bibitem[McKenzie \& V\"olk(1982)]{VK}McKenzie, J.F., V\"olk H.J., 1982, Astron. \& Astrophys., 116, 191.
\bibitem[Marcowith {\it et al.}(2006)]{MPL05}Marcowith, A., Lemoine, M., Pelletier, G., 2006, Astron. \& Astrophys., accepted (paper II).
\bibitem[Maron \& Goldreich(2001)]{MG}Maron J., Goldreich P., 2001, ApJ 554, 1175.
\bibitem[Melrose(1986)]{MEL} Melrose D.B., 1986, ``Instabilities in space and
laboratory plasmas", Cambridge University Press.
\bibitem[Moffat(1978)]{MOF} Moffatt H.K., 1978, ``Magnetic filed generation in
electrically conducting fluids", Cambridge University Press.
\bibitem[Parizot et al. (2006)]{Petal05} Parizot, E., Marcowith, A., Ballet, J. \& Gallant, Y.A., Astron. \& Astrophys. , 2006, submitted
\bibitem[Parizot et al. (2004)]{Petal04} Parizot, E., Marcowith, A., van der Swaluw, E., Bykov, A.M. \& Tatischeff, V., 2004, Astron.\& Astrophys., 424, 747
\bibitem[Pelletier \& Kersal\'e(2000)]{PK}Pelletier G., Kersal\'e E., 2000, Astron. \& Astrophys. , 361, 788.
\bibitem[Pohl et al (2005)]{Pohletal05} Pohl, M., Yan, H. \& Lazarian, A., 2005, \apj, 626, L101.  
\bibitem[Ptuskin \& Zirakshvili(2003)]{PT}Ptuskin V.S., Zirakashvili V.N., 2003, Astron. \& Astrophys., 410, 189.
\bibitem[Ptuskin \& Zirakshvili(2005)]{PZ05}Ptuskin V.S., Zirakashvili V.N., 2005, Astron. \& Astrophys., 429, 755.
\bibitem[Schlickeiser(1998)]{SC}Schlickheiser R.,Vainio R., 1998, Astron. \& Astrophys., 331, 793.
\bibitem[Skilling(1975)]{SKI}Skilling J., 1975, MNRAS, 172, 557; 173, 245.
\bibitem[Vink (2004)]{V04} Vink, J. 2004, Adv. Space Sci., 33, 356.
\bibitem[V\"olk et al.(2005)]{Vetal05}  V\"olk H.J., Berezhko, E. G., Ksenofontov, L. T., Astron. \& Astrophys., 433, 229.
\bibitem[Wentzel(1969)]{WEN}Wentzel D.G., 1969, Astrophys. J., 156, 303. 
\end{thebibliography}
\end{document}